# Microeconomics of nitrogen fertilization in boreal carbon forestry

Petri P. Kärenlampi

Lehtoi Research, Finland
petri.karenlampi@professori.fi

## Abstract

Nitrogen fertilization of boreal forests is investigated in terms of microeconomics, as a tool for carbon sequestration. The effects of nitrogen fertilization's timing on the return rate on capital and the expected value of the timber stock are investigated within a set of semi-fertile, spruce-dominated boreal stands, using an inventory-based growth model. Early fertilization tends to shorten rotations, reducing timber stock and carbon storage. The same applies to fertilization after the second thinning. Fertilization applied ten years before stand maturity is profitable and increases the timber stock, but the latter effect is small. Fertilization of mature stands, extending any rotation by ten years, effectively increases the carbon stock. Profitability varies but is increased by fertilization, instead of merely extending the rotation.

## Introduction

Growing forests sequester atmospheric carbon dioxide, partially mitigating the ongoing change of Earth's climate [1,2,3]. Enhancing the rate of forest growth advances the mitigation process. Various tools have been proposed for increasing the growth rate, such as plant improvement, ditch maintenance, thinning schedule optimization, and fertilization [4,5,6].

Nitrogen fertilization appears to be most effective when combined with the addition of phosphorus [7]. Nitrogen fertilization is not very beneficial on the most fertile sites [8,9,10]. Nitrogen fertilization mostly applies to mineral soil [11,12]. The trees on the site must be both abundant and vital enough to utilize the fertilization effect [8,9]. Conifers respond well to fertilization, whereas birch (*Betula*) species are less responsive [13,14,12,15].

As the effect of boron fertilization may endure [16], nitrogen fertilization usually contributes for a period of at most 10 years [7,17]. Long-term nitrogen fertilization may reduce microbial activity in soil, possibly resulting in increased soil organic matter content [18,12].

Arithmetically, one cubic meter of roundwood stores carbon the equivalent of one ton of carbon dioxide. However, carbon is also stored in roots, branches, leaves, litter, and soil. Altogether, the carbon storage of one cubic meter of commercial wood corresponds to about two tons of stored $CO_2$ [19,20].

An earlier paper has discussed the timing and intensity of fertilization in terms of a cash flow analysis [9]. A financial perspective is here adopted, instead of the cash-flow approach, and the carbon storage aspect is considered, in addition to business economics. Using an inventory-based growth model on five example stands, four different timings of nitrogen fertilization within any rotation are examined, discussing their effect on the rate of return on capital on the one hand, and on the expected value of the timber stock on the other.

## Materials and methods

Within a cyclical system, with periodic boundary conditions, the expected value of the profit rate is

$$\left\langle \frac{d\kappa}{dt} \right\rangle = \int_b^{b+\tau} \frac{d\kappa}{dt} \ p(t) \ dt \qquad (1),$$

where $\tau$ is cycle duration, $p(t)$ is the probability density of time within the cycle, and $\frac{d\kappa}{dt}$ is any current profit rate. On the profit/loss – basis, the profit rate includes value growth, operative expenses, interests, and amortizations, but neglects investments and withdrawals. On the other hand, the expected value of the capitalization is

$$\left\langle K \right\rangle = \int_b^{b+\tau} K \ p(t) \ dt \qquad (2),$$

where the capitalization $K$, on the balance-sheet basis, is directly affected by any investment and withdrawal. Then, the expected value of the rate of return on capital is

$$\left\langle r \right\rangle = \frac{\left\langle \frac{d\kappa}{dt} \right\rangle}{\left\langle K \right\rangle} \qquad (3).$$

For the application of Eq. (3) in rotation forestry, it is necessary to include growth rate, prices, and expenses. It also is necessary to include some kind of initial stand conditions, which may consist of either establishment procedures of a stand of seedlings or saplings, or measurement data of young, preferably not previously thinned, stands [21,22]. Here, measurement data from five never-thinned Norway spruce (*Picea abies*) -dominated young stands are taken as the set of initial stand conditions. The stands, of age between 30 and 45 years, have been described in detail in earlier papers [23,24,25,26,21,22]. This study discussing nitrogen fertilization, only mesic stands (medium fertility) are included, excluding herb-rich stands where nitrogen fertilization does not apply well [8,9,10].

Prices and expenses are here retained at the 2019 level, to retain comparability with earlier investigations [23,24,25,26,21,22]. Regeneration expenses are amortized first at the occurrence of final harvesting [27], whereas fertilization expenses are amortized at the occurrence of the first harvesting after fertilization.

Technically, a time evolution from any initial condition is established according to a growth model [28], in terms of 30-month timesteps. In the absence of thinnings, this procedure results in an expected value of capital return rate for any rotation age τ, according to Eq. (3), observable at the end of any time step. The rotation age giving the greatest expected value of the capital return rate appears the most feasible, in the absence of thinnings. Then, one thinning is introduced, experimenting with its timing, severity, allocation to tree species and diameter classes, within the computer program. If the thinning is successful in improving the maximal expected value of capital return rate, it is considered feasible. If thinning succeeds in improving the expected value of the capital return rate, another thinning is introduced, again experimenting with timing, severity, and allocation, now regarding both introduced thinnings. Further thinnings are introduced this way, one by one, provided the previous one is successful in improving the expected value of the capital return rate. It is worth noting that in principle, two or more thinnings could be financially feasible even if a single commercial thing would not be profitable. However, the author is not aware of any such occurrence in boreal forestry, where the number of thinnings tends to be limited because of the requirements of operational efficiency.

The above procedure did not contain any fertilization treatments. Four different sets of boundary conditions for fertilizations were introduced as follows.

As the five example stands have been observed at the age of 30…45 years, stem count 1655…2451/ha, and basal area 29…49 $m^2$/ha, they were due for commercial thinning. The first fertilization treatment examined was implemented without delay after the thinning. Two of the five example stands appearing in were supposed to be thinned twice, to maximize the expected value of the return rate on capital (Eq. (3)). The other set of boundary conditions was to fertilize after the second thinning. While early fertilizations possibly shorten rotations, the third boundary condition was fertilization ten years before maturity, supposedly not shortening rotations. The fourth boundary condition was to fertilize at stand maturity, which probably would extend any rotation by ten years.

The effect of any nitrogen fertilization was technically implemented within the growth model [28] by increasing the site fertility index by five meters (dominant height at breast-height age of 40 years) for a period of ten years. It was verified that the growth response gained this way did correspond with the results of experimental fertilization studies [12,7,15,8,9].

**Results**

As the five example stands have been observed at the age of 30…45 years, stem count 1655…2451/ha, and basal area 29…49 $m^2$/ha, they were due for commercial thinning. The first fertilization treatment examined was implemented without delay after the thinning. The results, regarding the expected value of the operative rate of return on capital, are shown in Fig. 1. For comparison, Fig. 1 also shows the corresponding results in the absence of fertilization. It is found that in all five cases, fertilization increases the return rate on capital. On the other hand, fertilization reduces the rotation age, corresponding to the rotation resulting in the greatest return rate on capital. The shortening of the rotation was five years in four cases and ten years in one case.

As the rotations are shortened, the expected value of the timber stock is reduced, as shown in Fig. 2, with one exception. The reduction of the timber stock is in the order of 5 to 10 $m^3$/ha.

It is worth noting that the expected value of the timber stock is elevated already at the leftmost data points appearing in Fig. 1. These data points correspond to a time spot five years after the initial field observation. The observable difference in the expected value of the return rate on capital is affected by the fact that the suitable thinning intensity is not the same if the stand is supposed to be fertilized; application of the fertilization favors a greater timber stock. Such a phenomenon is most pronounced in the case of stand 14/06, where the expected value of the timber stock increases according to Fig. 2.

In the presence of fertilization, the operative return rate on capital is maximized if the rotations are shortened in Fig. 1. However, there is no obligation for shortening rotations. In four of the five cases, fertilization would be slightly profitable even if the rotations would not be shortened (Fig. 1). Duration of the rotation could be contracted in a carbon sequestration agreement. However, in Fig. 1, in the absence of fertilization, the maximum return rate on capital is gained by final harvesting from 15 to 40 years after fertilization. It might be challenging to regulate rotation ages in such a long time frame.

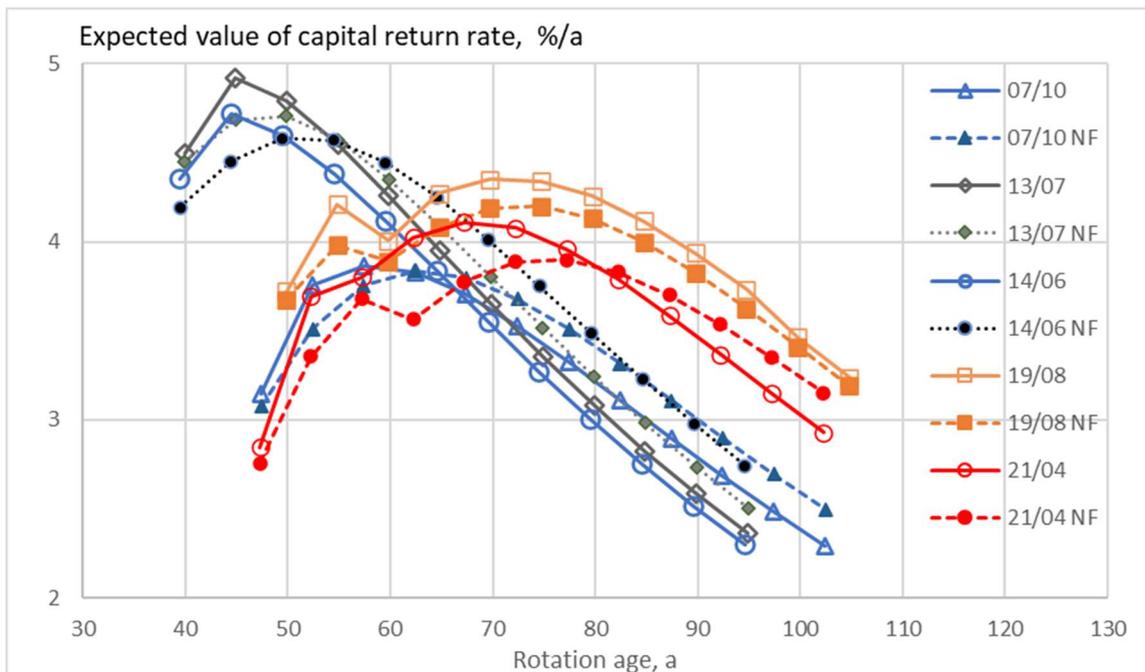

Figure 1. Expected value of capital return rate with fertilization applied after the first thinning, and without fertilization.

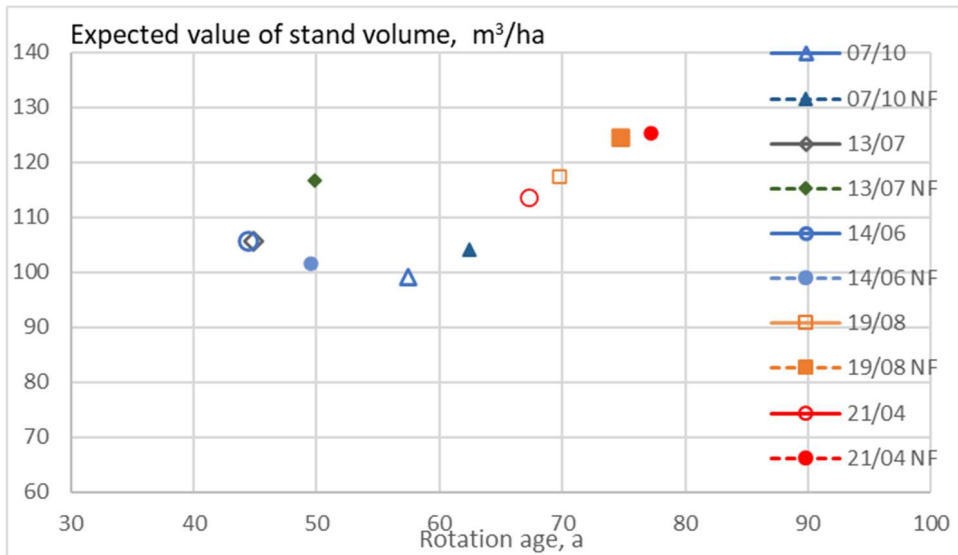

Figure 2. Expected value of stand volume, with fertilization after the first thinning, and without fertilization.

Two of the five example stands appearing in Fig. 1 are supposed to be thinned twice, to maximize the expected value of the return rate on capital (Eq. (3)). Then, a natural alternative for fertilization after the first thinning, shown in Fig. 1, would be fertilization after the second thinning. The result, in terms of the expected value of the return rate on capital, is shown in Fig. 3. In both cases, fertilization is profitable. However, not only the return rate on capital is increased, but the optimal rotation is shortened by ten years in both cases. In case the rotation would not be shortened, the fertilization would be non-profitable, as it would reduce the return rate on capital (Fig. 3).

As the rotation age is shortened by a decade, the expected value of the timber stock is reduced, as shown in Fig. 4. This is even though forthcoming fertilization induces an elevated timber stock to be retained after the second thinning.

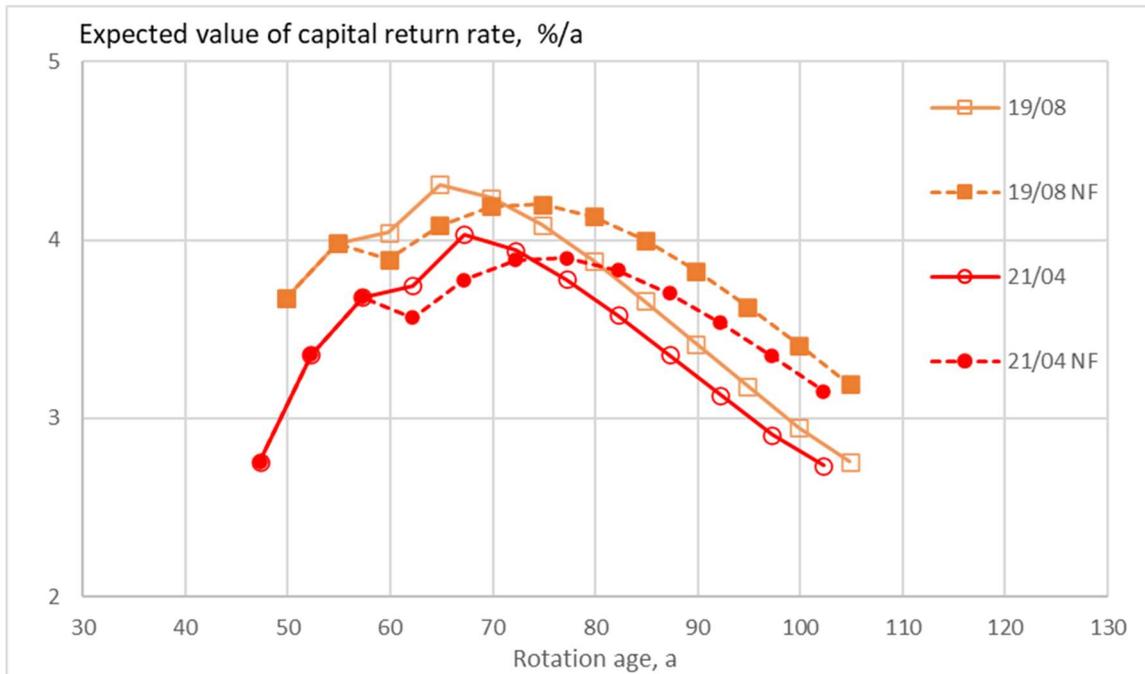

Figure 3. Expected value of capital return rate with fertilization applied after the second thinning, and without fertilization.

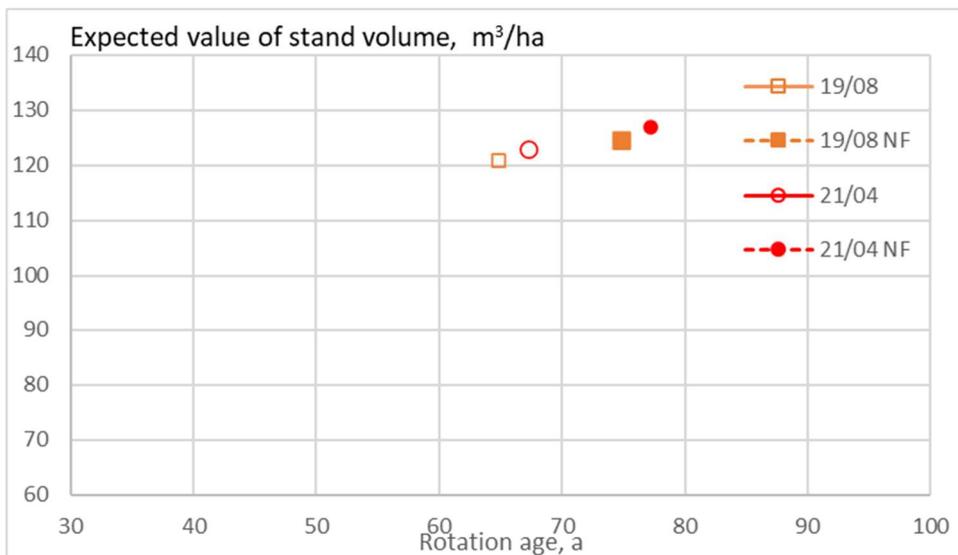

Figure 4. Expected value of stand volume, with fertilization after the second thinning, and without fertilization.

Obviously, Figures 1 to 4 indicate that early fertilization is not necessarily a functional tool of carbon sequestration, as it tends to shorten rotations and correspondingly reduce the expected value of the timber stock. An obvious solution for the problem is that fertilization is delayed until ten years before stand maturity, in which case it should not shorten rotations. Figure 5 shows the results of such a procedure. It is found that in all five cases, fertilization is profitable, as it increases the expected value of the return rate on capital.

Figure 6 shows the expected value of the timber stock, with and without fertilization. In all cases, fertilization increases the expected value of the timber stock. However, in all cases, the effect is small, in the order of one to three percent. The small effect is understandable since the timber stock is slightly elevated for a small fraction of the rotation [29]. It is worth noting that as the fertilizations in Figs. 5 and 6 are not associated with any thinning, the thinning schedules have not been altered.

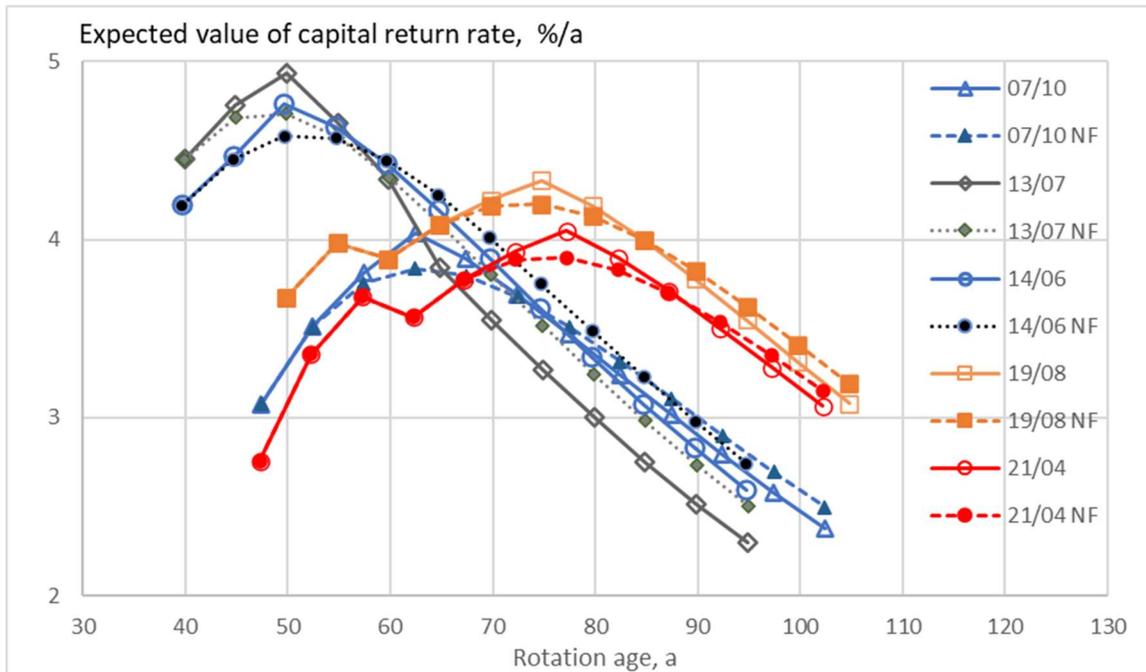

Figure 5. Expected value of capital return rate with fertilization applied ten years before stand maturity, not altering the rotation time, and without fertilization.

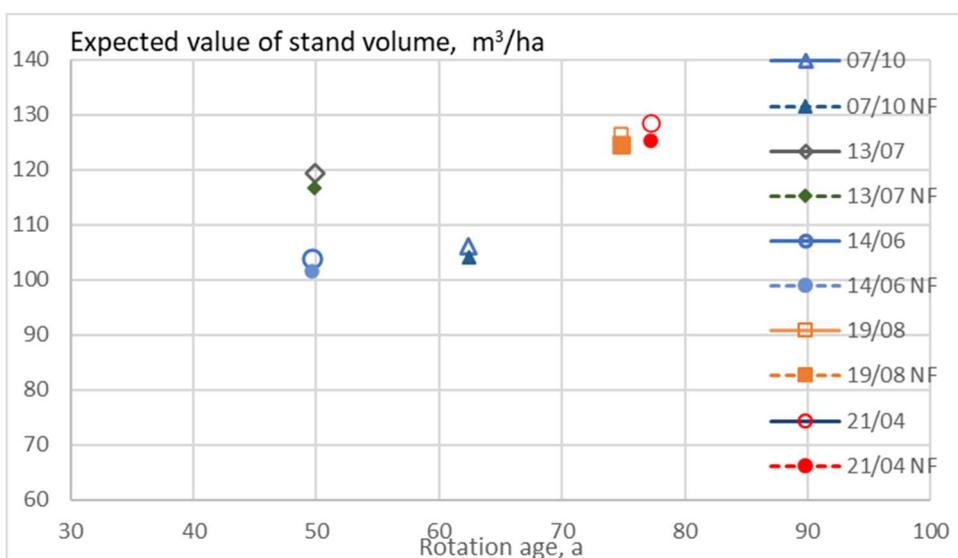

Figure 6. Expected value of stand volume, with fertilization ten years before stand maturity, not altering the rotation time, and without fertilization.

Obviously, a larger increment of the expected value of the timber stock would be gained if the rotations would be extended. Figure 7 shows the expected value of the return rate on capital when fertilization is applied at stand maturity, extending any rotation by ten years. Profitability varies: in two of the five cases the treatment is profitable, whereas in the case of three cases, it is not. However, fertilization improves the profitability in all cases, in comparison to merely extending the rotation (Fig. 7).

Figure 8 shows that the extension of the rotation by ten years is an effective tool for increasing the timber stock. In the presence of fertilization, the expected value of the timber stock is increased from 17.8% to 29.1%, the smallest increments gained by extending the longest rotations. In terms of cubic meters per hectare, these values correspond to 22…34 m$^3$/ha. In the absence of fertilization, the timber stock increment was from 16.6% to 26.9%, or 21…31 m$^3$/ha. Correspondingly, the fertilization effect on the timber stock increment is small. It appears from 7 that the financial effect probably is more important.

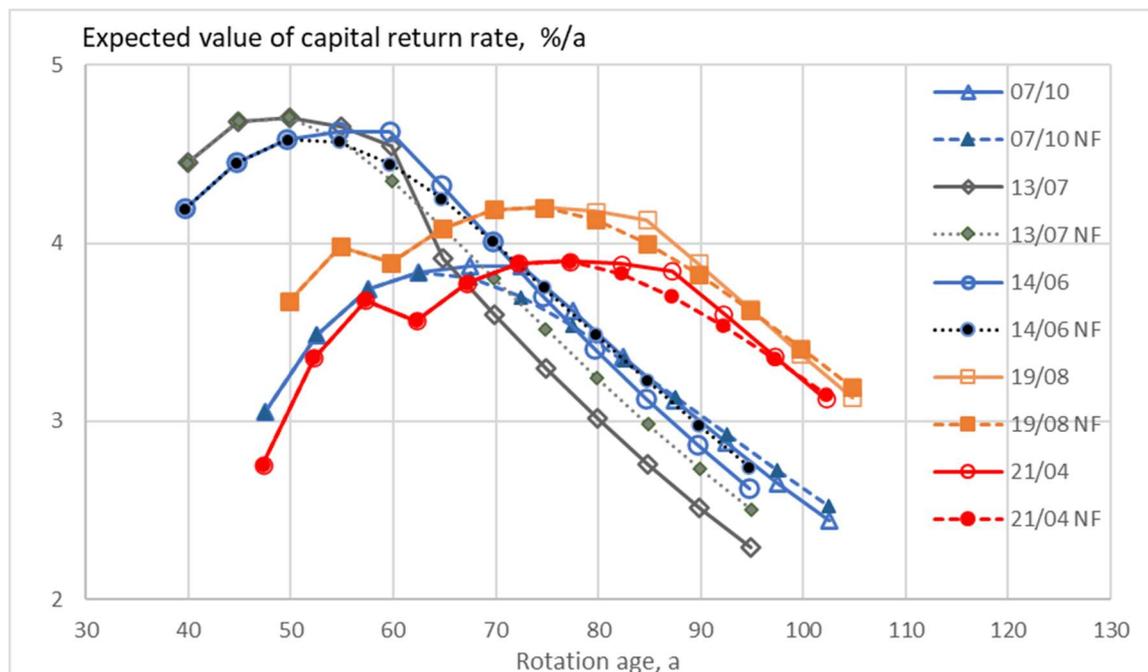

Figure 7. Expected value of capital return rate with fertilization applied at stand maturity, prolonging the rotation by at least 10 years, and without fertilization.

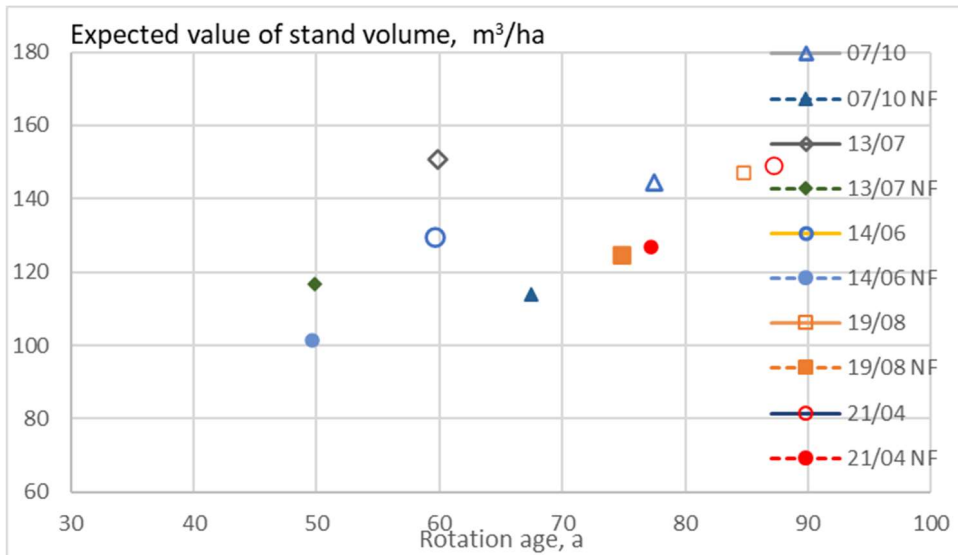

Figure 8. Expected value of stand volume, with fertilization at stand maturity, inducing rotation extension of ten years, and without fertilization.

It appears possible to quantify the financial effect of extending the rotation, as well as the effect of fertilization on it. The financial expense of extending the rotation is

$$\langle E \rangle = -\Delta \langle r \rangle \; (\tau + \Delta \tau) \; (\langle C \rangle + \Delta \langle C \rangle) \quad (4),$$

where $\Delta \langle r \rangle$ refers to the change in the expected value of the rate of return on capital, $\Delta \tau$ to the change in the duration of the rotation, and $\Delta \langle C \rangle$ to the change in the expected value of capitalization.

The financial expense of enhanced timber stock per year and excess volume can in turn be given as

$$\frac{\langle E \rangle}{\Delta \langle V \rangle \; (\tau + \Delta \tau)} \qquad (5),$$

where $\Delta \langle V \rangle$ is the change in the expected value of commercial stand volume.

Figure 9 shows that the financial expense of extending the rotation by ten years, according to Eq. (4), varies from 548 to 1628 Eur/ha without fertilization. Fertilization significantly reduces the financial expense, the rotation extension becoming profitable in the case of two of the five example stands. It is still worth noting that the expenses are on the level of year 2019 [23,24,27].

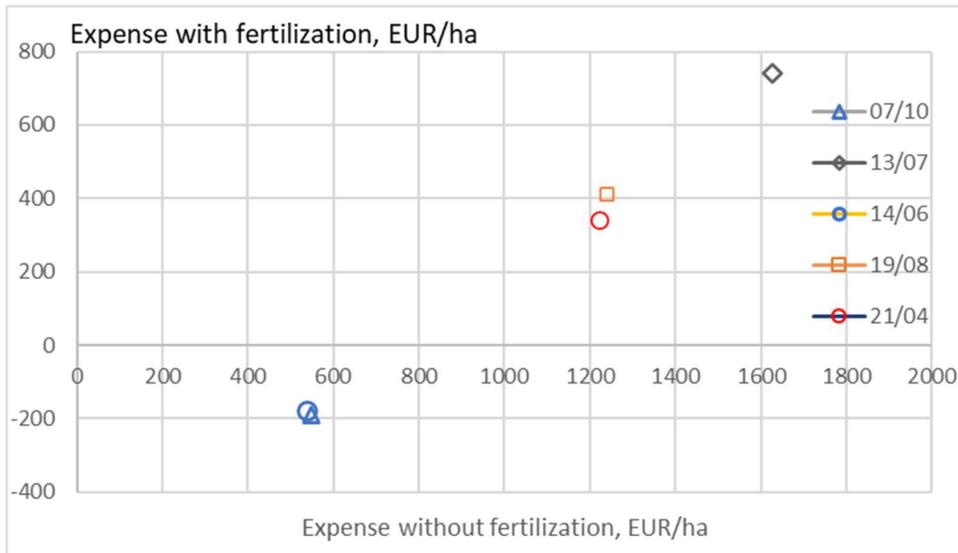

Figure 9. The financial expense of extending the rotation by ten years, according to Eq. (4). Without fertilization (horizontal axis) and with fertilization (vertical axis).

It is further possible to compute the financial expense of enhanced timber stock per year and excess volume according to Eq. (5). Without fertilization, the financial expense is 0.32…0.80 Eur/(m³*a), and with fertilization -0.10…0.36 Eur/(m³*a). From Eq. (5), the financial expenses allocated for the extension period only are achievable by scaling with $\left(\tau + \Delta\tau\right)/\Delta\tau$.

**Discussion**

The results indicate that early fertilization tends to shorten rotations, and thereby reduce the expected value of the timber stock volume. Correspondingly, early fertilization cannot be used as a tool for carbon sequestration. There may be circumstances where this conclusion does not hold. In the case of a centrally planned economy, instead of a market economy with microeconomic drivers, it might be possible to regulate the rotation ages. Even in a market economy, it might be possible to regulate the rotation ages in terms of voluntary carbon sequestration contracts. However, difficulties might appear in the realization of long-term contracts, applicable for individual forest sites for several decades.

The results further indicate that as fertilization implemented first ten years before stand maturity does not shorten rotations, it correspondingly does not reduce the carbon stock. Such an operation is often profitable (Fig. 5), but the effect on carbon storage is small (Fig. 6). The small contribution to carbon storage has previously been discussed in the literature [29].

Fertilization applied at stand maturity, thereby extending rotations, effectively increase carbon storage (Fig. 8). However, as indicated in the text, the effect of the fertilization on the increment of the expected value of the timber stock is small, and the main contribution comes from the extension of the rotation as such. On the other hand, the effect of fertilization on finances is large, as indicated in Fig. 9. Fertilization significantly increases profitability according to Fig. 9, but Fig. 7 indicates that even if fertilization is applied, the extension of the rotation is mostly not profitable. Thus, microeconomics proposes that the extension of rotations requires some kind of external incentive: a carbon sequestration contract, or the like.

There are circumstances where the results of this paper do not apply. Firstly, nitrogen fertilization is not very beneficial on the most fertile sites [8,9,10] – such sites have been excluded from the present dataset. Secondly, nitrogen fertilization only applies to mineral soil [11,12]. Thirdly, the trees on the site must be both abundant and vital enough to utilize the fertilization effect [8,9].

The denominator of Eq. (3) naturally includes bare land value. Bare forest land is not frequently traded, which makes it difficult to assess its market value. The value of bare land does contribute to the results: increased bare land value favors greater timber stocks and longer rotations. The effect of variable bare land value on fertilization has not been investigated in this study.

Instead of the financial formulation given in Eq. (3), the present problematics could have been approached in terms of cash flow analysis [9]. Such of an approach was recently attempted also by this author. Several problems appeared. Firstly, the overall results are very sensitive to discount rates. Secondly, the effect of temporal displacement of activities is sensitive to discount rates. Thirdly, it is difficult to compare the current profitability of fertilizations implemented at different times. Fourthly, the profitability of fertilizations implemented at different times compared at any observation time (present time) depends on the selection of the observation time.

There may be other circumstances where the microeconomic approach applied in this paper is not functional. The present approach is based on the computation of an operative return rate of capital, under periodic boundary conditions. The periodic boundary conditions as such are a simplification of reality – however, not necessarily unrealistic. More importantly, the periodic boundary condition possibly can be abandoned by divesting estates [30]. Then, real estate proceeds may complement or exceed the income from forest operations. This might open new avenues for the utilization of early fertilization. Any more detailed analysis of such economics however is outside the scope of this study.

This paper has discussed the microeconomics of fertilization and the timing of harvesting, along with its consequences in carbon sequestration. The outcome, however, may depend on macroeconomic boundary conditions [6]. An extreme macroeconomic boundary condition might be a stiff (constant) demand for timber. Such a boundary condition would result in reduced harvesting elsewhere if harvesting would be increased on one estate. If increased growth does not increase harvesting, increased growth always is beneficial for the accumulation of the carbon stock. However, the present author is not aware of any proof that such stiff boundary conditions in demand would exist. Instead, one can reasonably assume that macroeconomics is a system producible by integrating microeconomies, even if interactions between microeconomies are difficult to predict [31, 32, 33, 34].

## Acknowledgement

This research was partially financed by Niemi-foundation.